\documentclass{webofc}
\usepackage[varg]{txfonts}   % Web of Conferences font
\usepackage{listings}
\usepackage{hyperref}
\begin{document}
\title{How Rotation Affects Masses and Ages of Classical Cepheids}


\author{\firstname{Richard~I.} \lastname{Anderson}\inst{1}\fnsep\thanks{\href{mailto:ria@jhu.edu}{\tt ria@jhu.edu}},
  \firstname{Sylvia} \lastname{Ekstr\"om}\inst{2},
  \firstname{Cyril} \lastname{Georgy}\inst{2},
  \firstname{Georges} \lastname{Meynet}\inst{2},
  \and
  \firstname{Hideyuki} \lastname{Saio}\inst{3}
}

\institute{Department of Physics and Astronomy, The Johns Hopkins University, Baltimore, MD 21218, USA
\and
           D\'epartment d'Astronomie, Universit\'e de Gen\`eve, 1290 Sauverny, Switzerland
\and
           Astronomical Institute, Graduate School of Science, Tohoku University, Sendai, 980-8578 Miyagi, Japan
          }

\abstract{%
Classical Cepheid variable stars are both sensitive astrophysical laboratories and accurate cosmic distance tracers. We have recently investigated how the evolutionary effects of rotation impact the properties of these important stars and here provide an accessible overview of some key elements as well as two important consequences. Firstly, rotation resolves the long-standing Cepheid mass discrepancy problem. Second, rotation increases main sequence lifetimes, i.e, Cepheids are approximately twice as old as previously thought. Finally, we highlight the importance of the short-period ends of Cepheid period distributions as indicators for model adequacy.
}
\maketitle

\section{Introduction}\label{sec:intro}

Classical Cepheid variable stars (henceforth: Cepheids) are yellow (super-)giant stars of intermediate mass ($3 - 10 M_\odot$) well-suited for investigating how rotation affects stellar evolution in general. The evolutionary status of Cepheids can be accurately inferred, since these stars occupy a well-defined region in the Hertzsprung-Russell-Diagram, the classical instability strip (IS). Additionally, Cepheid variability is very noticeable and straightforward to identify. Due to the timescales involved, nearly all Cepheids are observed during the blue loop evolutionary phase, i.e., during second or third IS crossings, which are inferred using observed rates of period change (\cite{2006PASP..118..410T}). Since blue loop evolution is highly sensitive to model assumptions and input physics (\cite{Kip+Weig}), the same holds for Cepheid properties.

The impact of rotation on stellar evolution increases with mass (\cite{2000ARA&A..38..143M}), and Cepheids occupy somewhat of a ``sweet spot'' in this regard. They are massive enough for rotational mixing to be relevant, but not as dominated by binary interactions and radiatively-driven winds as more massive stars (\cite{2012Sci...337..444S,2008A&ARv..16..209P}). That said, many Cepheids are part of binary systems (\cite{2003IBVS.5394....1S,2015AJ....150...13E,2016ApJS..226...18A}) and interactions may occur during various stages of a star's evolution (\cite{2015A&A...574A...2N,2015ApJ...804..144A}).

Geneva stellar evolution models provide insights into the effects of rotation for a wide range of masses and metallicities \cite{2012A&A...537A.146E,2013A&A...553A..24G}. Capitalizing on their sensitivity to stellar physics, we have carried out the first detailed investigations of how rotation impacts the evolution and pulsation properties of Cepheids (\cite{2014A&A...564A.100A,2016A&A...591A...8A}). In this contribution, we provide an accessible overview of some key notions regarding rotation and argue that rotation is required to correctly predict Cepheid properties. Specifically, we focus on the Cepheid mass discrepancy (\S\ref{sec:mass}) and period-age relations (\S\ref{sec:P-A}). We close by highlighting the importance of correctly reproducing the short-period end of the Cepheid period distribution as well as the need to test predictions that are unmistakably linked to rotation (\S\ref{sec:disc}).

\begin{figure}
\centering
\sidecaption
\includegraphics{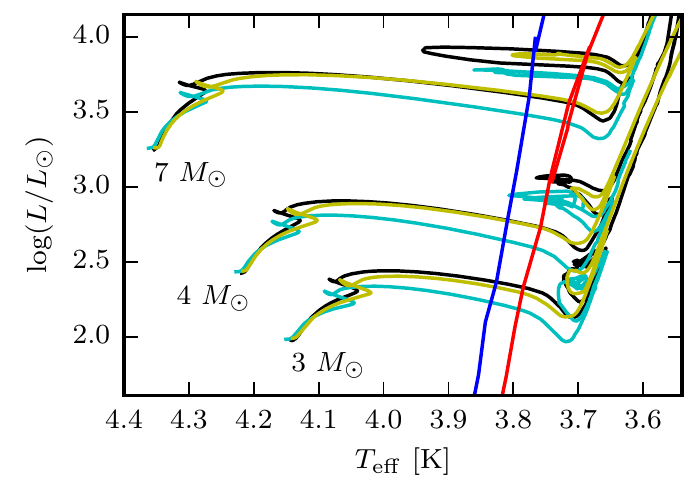}
\caption{Evolutionary tracks for 3, 4, and 7 $M_\odot$ models with LMC metallicity (Z=0.006). Cyan and black lines: Geneva models without and with average initial (ZAMS) rotation ($\omega=\Omega/\Omega_{\rm{crit}}=0.5$). Yellow lines: MESA models (\cite{2013ApJS..208....4P}) with $v/v_{\rm{crit}} = 0.4$, i.e., $\omega=0.586$. $5\,M_\odot$ models not shown for clarity. Blue and red IS boundaries from \cite{2016A&A...591A...8A}. Cepheids are highly sensitive to model assumptions and input physics, being observed almost exclusively during blue loop evolution, i.e., $2^{\rm{nd}}$ and $3^{\rm{rd}}$ IS crossings. Rotation modifies the mass-luminosity relation and resolves the long-standing Cepheid mass discrepancy (\S\ref{sec:mass}).}
  \label{fig:tracks}
\end{figure}

\section{How rotation resolves the Cepheid mass discrepancy problem}\label{sec:mass}
\begin{figure}[b]
  \centering
  \sidecaption
  \includegraphics{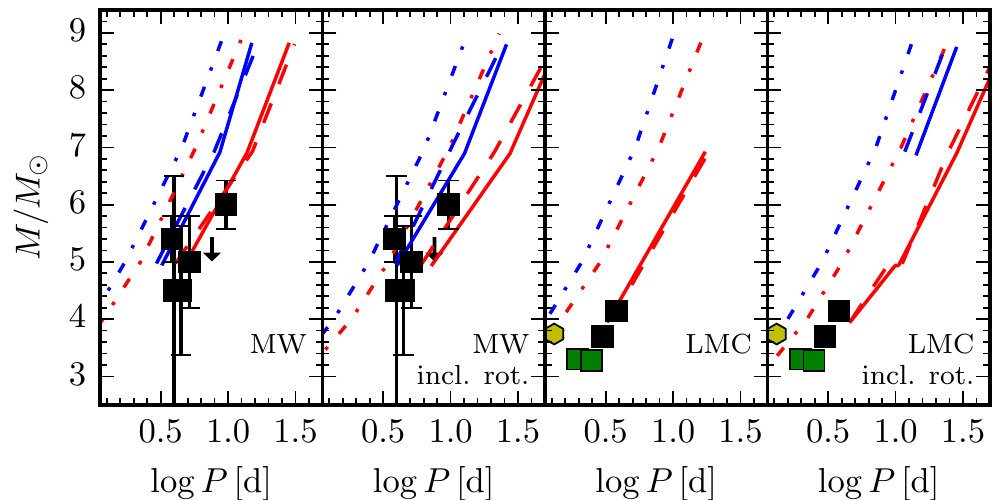}
  \caption{Fundamental mode mass-period relations for Solar (Z=0.014, MW) and LMC (Z=0.006) models with and without rotation compared to model- independent mass estimates for fundamental mode (black squares), first overtone (green squares), and one anomalous Cepheid (yellow hexagon), cf. \cite{2016A&A...591A...8A} and references therein.}
  \label{fig:MLR}
\end{figure}

Much work has been done to resolve the long-standing Cepheid mass discrepancy problem (cf. \cite{2008ApJ...677..483K,2006MmSAI..77..207B}, and references therein), which manifests in a $10-20\%$ overestimate of Cepheid masses inferred from stellar evolution models compared to other (pulsational, or model-independent) mass estimates. Conceptually, there are two possibilities of resolving the mass discrepancy: a) lowering the mass of a given model at fixed luminosity, e.g. via enhanced mass-loss (\cite{2008ApJ...684..569N}), and b) increasing luminosity at fixed mass, e.g. by increasing effective stellar core sizes via enhanced convective core overshooting (e.g. \cite{2012ApJ...749..108P}) or rotational mixing (\cite{2014A&A...564A.100A}). Despite growing observational evidence for circumstellar material in the vicinity of Cepheids \cite{2010ApJ...725.2392M,2009A&A...498..425K,2017A&A...597A..73N}, pulsationally-enhanced mass-loss rates (\cite{2008ApJ...684..569N}, $\dot{M} \sim 10^{-10} - 10^{-8} M_\odot \rm{yr}^{-1}$) do not appear sufficient to resolve the mass discrepancy given that the Cepheid evolutionary phase is relatively short-lived ($10^4 - 10^6$\,yr). Rotation, however, provides a sufficient increase in luminosity to resolve the mass discrepancy, cf. Fig.~\ref{fig:tracks} and \ref{fig:MLR}.

Geneva stellar evolution models include convective core overshooting ($d_{\rm{over}}/H_P = 0.1$) to match the observed width of the main sequence over the mass range $1.35-9\,M_\odot$ (\cite{2012A&A...537A.146E}) and have not been adjusted to reproduce properties of evolved stars. Notably, Geneva models include a consistent treatment of (radially) differential rotation that induces mixing throughout a star at all times during its evolution. Rotational mixing occurs even in radiative layers, notably replenishing the core's hydrogen supply while bringing processed materials to the surface during main sequence evolution. Thus, rotational mixing increases the core's effective mass, although hydrostatic effects (centrifugal forces) counteract this increase for very fast rotation (\cite{2014A&A...564A.100A}). Due to this competition, luminosity does not increase monotonously with increasing rotation. Rather, luminosity reaches a maximum near typical initial angular rotation rates, varies relatively slowly over the range of most probable $\omega = \Omega/\Omega_{\rm{crit}} \sim 0.5$, and falls off towards both extremes. However, even the models with the fastest rotation feature higher luminosities than models with no rotation.

Rotation thus modifies the mass-luminosity relation (MLR). Since initial angular velocities ($\omega$) differ from star to star, it follows that MLRs have intrinsic scatter. Of course, MLRs also change during the evolution of a star, e.g. between $2^{\rm{nd}}$ and $3^{\rm{rd}}$ IS crossings (Fig.~\ref{fig:tracks}). Hence, to resolve the mass discrepancy, both effects (rotation and crossing) must be considered simultaneously (\cite{2014A&A...564A.100A}). Mass-period relations provide  more direct observational tests, since pulsation periods are easily measured and since luminosity differences between $2^{\rm{nd}}$ and $3^{\rm{rd}}$ crossings imply changes in stellar structure that affect the pulsation equation ($P = Q/\sqrt{\bar{\rho}}$), where $\bar{\rho}$ denotes average density and $Q$ is the pulsation ``constant'' (\cite{1998ApJ...498..360S}). Figure~\ref{fig:MLR} shows that observed and predicted mass-period relations in the Milky Way and Large Magellanic Cloud (LMC) agree for models that include average levels of rotation.

The intrinsic scatter of MLRs implied by rotation is further relevant for predicting the period-luminosity-color (PLC) relation, which follows from inserting an \textit{assumed} MLR into the pulsation equation combined with the Stefan-Boltzmann law. Given that the MLR itself is not unique, the PLC-relation cannot be unique, either. It would be interesting to test the impact of any associated effects (e.g. changes in He abundance) on Cepheid light and radial velocity variations (\cite{2013ApJ...768L...6M}).

\section{Rotational rejuvenation: Cepheids older than previously thought}\label{sec:P-A}
\begin{figure}[b]
  \centering
\sidecaption
\includegraphics{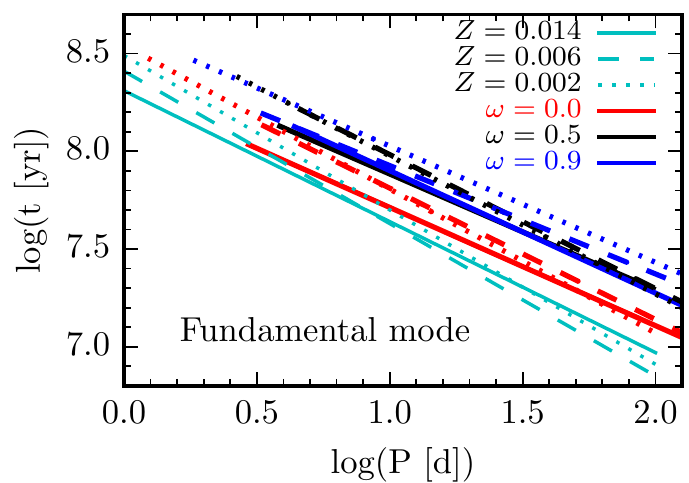}
\caption{Period-age relations for fundamental mode Cepheids \cite{2016A&A...591A...8A}. Linestyles differentiate Solar, LMC, and SMC metallicity and colors indicate ZAMS rotation rate ($\omega$) as indicated. Cyan lines are literature period-age relations that do not take into account rotation and are on a slightly different metallicity scale ($Z_\odot=0.02$, $Z_{\rm{LMC}}=0.008$, $Z_{\rm{SMC}}=0.004$, \cite{2005ApJ...621..966B}). Rotation extends main sequence lifetimes by mixing unprocessed material into the core during H-burning, resulting in significantly ($\Delta \log{t} = 0.2 - 0.3$\,dex) older Cepheids.}
\label{fig:period-age}       % Give a unique label
\end{figure}

Period-age relations (\cite{1969A&A.....1..142K}) provide useful age estimates of stellar populations based on readily-observed pulsation periods. The age of a star in the Cepheid stage is dominated by its main sequence lifetime, followed by the duration of core He burning before the onset of blue loop evolution and the evolution along it.
Rotation significantly prolongs the duration over which a star can produce energy via fusion in its core by effectively increasing core size, leading to increased lifetimes for increasing levels of rotation. Note that period-age relations for models with average and fast rotation are comparable, in contrast to the effects of rotation on MLRs (\S\ref{sec:mass}, \cite{2014A&A...564A.100A}). Using Geneva models (\cite{2016A&A...591A...8A}), we have quantified this effect and find that typical levels of rotation imply Cepheids to be older by approximately $60-100\%$ than if rotation is not taken into account, cf. Fig.~\ref{fig:period-age}.

This change in Cepheid age scale is particularly important if Cepheid ages are compared with other age indicators. To illustrate the impact, ages for three $\sim 20$\,d Cepheids located in the nuclear bulge of the Galaxy (\cite{2011Natur.477..188M}) have been estimated to be $25\pm5$\,Myr using solar-metallicity period-age relations that do not account for rotation (\cite{2005ApJ...621..966B}). Yet, period-age relations that take into account typical initial rotation rates imply typical ages of $50-52\,$Myr, depending on the IS crossing. In the future, it will be important to cross-check the Cepheid age scale with other (hopefully independent) age indicators, such as asteroseismic studies of lower-mass stars in which rotational mixing is less efficient. In particular, Cepheids located in open clusters (\cite{2013MNRAS.434.2238A}) are important test-beds for models, providing a means to compare period-age relations to cluster isochrone ages or potentially asteroseismic ages of cluster members. Further tests of the rotational mixing efficiency, especially via CNO surface abundances, will provide important cross-checks to evaluate the accuracy of the predictions.

\section{Conclusions}\label{sec:disc}

Rotation has significant implications for Cepheids, resolving the mass discrepancy problem and implying significantly older ages than previously thought. Rotation leads to additional testable predictions, since rotation impacts stellar structure, evolution, mass-loss, and surface abundances (\cite{2000ARA&A..38..143M}). To this end, we have carried out extensive comparisons with different kinds of observational data (\cite{2014A&A...564A.100A,2016A&A...591A...8A}) in the past and will continue to confront model predictions to observations.

\begin{figure}[b]
\centering
\sidecaption
\includegraphics{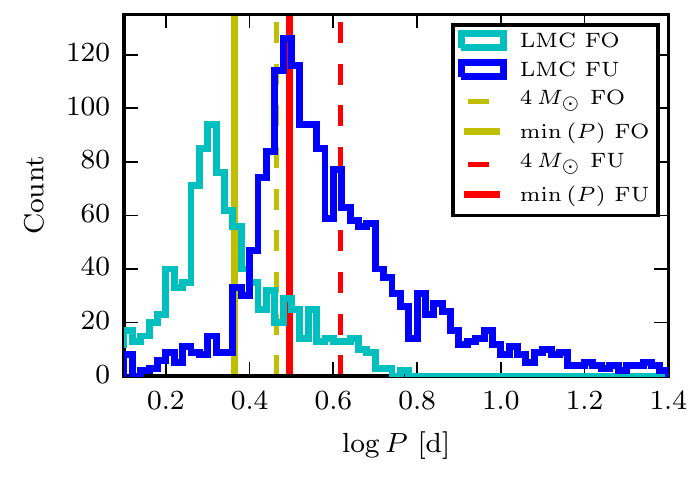}
\caption{Period distribution of OGLE fundamental mode (FU) and first overtone (FO) Cepheids (\cite{2008AcA....58..163S}) compared to periods predicted by Geneva models with Z=0.006 and average rotation (\cite{2016A&A...591A...8A}). Dashed vertical lines: lowest-mass ($4\,M_\odot$) computed model to enter IS. Solid vertical lines: grid-interpolated $3.7\,M_\odot$ model whose blue loop just barely reaches the red edge of the IS. Pulsation periods for interpolated models are based on a linear extrapolation of $\log{(\sqrt{\bar{\rho}})}-\log{(P)}$ relations (\cite{2016A&A...591A...8A}). The FU peak is well-matched by predictions. The puzzling mismatch for the FO peak requires further investigation.}
\label{fig:Pdist}       % Give a unique label
\end{figure}

A related key observable is \textit{the distribution of Cepheid pulsation periods}, which is sensitive to the blueward extent of blue loops: the lower the minimum mass of a model entering the IS, the lower the predicted minimum period. Short-period Cepheids are by far the most numerous in a population due to long IS crossing timescales and the shape of the initial mass function. Given that blue loop evolution is suppressed in models with high values of convective core overshooting, the short-period peak sets a stringent constraint on this important parameter. Fig.~\ref{fig:Pdist} shows that Geneva models correctly predict the peak in period for FU Cepheids\footnote{Minimum mass of models entering IS during blue loop evolution determined by \href{https://obswww.unige.ch/Recherche/evoldb/index/}{interpolation}}. Intriguingly, the predicted minimum period for FO Cepheids is longer than observed. These points will be investigated further using synthetic populations (\cite{2014A&A...566A..21G}).

CNO surface abundances probe the efficiency of main sequence rotational mixing, which is directly related to the Cepheid age scale. In a preliminary comparison using literature abundances, it appeared that CNO abundance surface enhancement is over-predicted by Geneva models \cite{RIAPHD}. However, rotation does appear to be required to correctly reproduce the extended main sequence turnoffs of intermediate-age open clusters \cite{2015ApJ...807...25B}. Further observational tests of these effects are required to improve the models and ensure the accuracy and consistency of stellar age scales.

The ESA space mission \textit{Gaia} will provide crucial information in this regard, allowing to firmly establish the often uncertain cluster membership of Cepheids (\cite{2013MNRAS.434.2238A}) and thus significantly improving the ability to use cluster populations  as test-benches for different assumptions, input physics, and numerical implementations required to model stellar evolution.

%% \begin{acknowledgement} 
%% \noindent\vskip 0.2cm
%% \noindent {\em Acknowledgments}: \textbf{Your acknowledgments, if any, should go in this field. }
%% \end{acknowledgement}

% BibTeX or Biber users please use (the style is already called in the class, ensure that the "woc.bst" style is in your local directory)
% \bibliography{name or your bibliography database}

\begin{thebibliography}{}
%%% To get the right format from ADS references, use as custom format:
%%% \\bibitem%{R}\n%\8.3l,%\q,\\textbf{%\V}, %\p (%y)\n

\bibitem{2006PASP..118..410T} 
Turner, D.~G., Abdel-Sabour Abdel-Latif, M., 
\& Berdnikov, L.~N., PASP, \textbf{118}, 410 (2006) 
  
\bibitem{Kip+Weig}
Kippenhahn, R. \& Weigert, A., \textit{Stellar Structure and Evolution} (Springer-Verlag, Berlin--Heidelberg--New York; also Astronomy and Astrophysics Library XVI, 1994)

\bibitem{2000ARA&A..38..143M} 
Maeder, A., \& Meynet, G., ARA\&A, \textbf{38}, 143 (2000) 

\bibitem{2008A&ARv..16..209P} 
Puls, J., Vink, J.~S., \& Najarro, F., A\&ARv, \textbf{16}, 209 (2008) 

\bibitem{2012Sci...337..444S} 
Sana, H., de Mink, S.~E., de Koter, A., et al., Sci, \textbf{337}, 444 
(2012) 

\bibitem{2003IBVS.5394....1S} 
Szabados, L., IBVS, \textbf{5394}, 1 (2003) 

\bibitem{2015AJ....150...13E} 
Evans, N.~R., Berdnikov, L., Lauer, J., et al., AJ, \textbf{150}, 13 (2015) 

\bibitem{2016ApJS..226...18A} 
Anderson, R.~I., Casertano, S., Riess, A.~G., et al., ApJS, \textbf{226}, 
18 (2016)

\bibitem{2015ApJ...804..144A} 
Anderson, R.~I., Sahlmann, J., Holl, B., Eyer, L., Palaversa, L., Mowlavi, 
N., S{\"u}veges, M., \& Roelens, M., ApJ, \textbf{804}, 144 (2015) 

\bibitem{2015A&A...574A...2N} 
Neilson, H.~R., Schneider, F.~R.~N., Izzard, R.~G., Evans, N.~R., 
\& Langer, N., A\&A, \textbf{574}, A2 (2015) 

\bibitem{2012A&A...537A.146E} 
Ekstr{\"o}m, S., Georgy, C., Eggenberger, P., et al., 
A\&A, \textbf{537}, A146 (2012) 

\bibitem{2013A&A...553A..24G} 
Georgy, C., Ekstr{\"o}m, S., Granada, A., Meynet, G., Mowlavi, N., 
Eggenberger, P., \& Maeder, A., A\&A, \textbf{553}, A24 (2013)

\bibitem{2014A&A...564A.100A} 
Anderson, R.~I., Ekstr{\"o}m, S., Georgy, C., Meynet, G., Mowlavi, N., 
\& Eyer, L., A\&A, \textbf{564}, A100 (2014) 

\bibitem{2016A&A...591A...8A} 
  Anderson, R.~I., Saio, H., Ekstr{\"o}m, S., Georgy, C., \& Meynet, G., A\&A, \textbf{591}, A8 (2016)

\bibitem{2013ApJS..208....4P} 
Paxton, B., Cantiello, M., Arras, P., et al., ApJS, \textbf{208}, 4 (2013) 

\bibitem{2006MmSAI..77..207B} 
Bono, G., Caputo, F., \& Castellani, V., MmSAI, \textbf{77}, 207 (2006) 

\bibitem{2008ApJ...677..483K} 
Keller, S.~C., ApJ, \textbf{677}, 483-487 (2008) 

\bibitem{2008ApJ...684..569N} 
Neilson, H.~R., \& Lester, J.~B., ApJ, \textbf{684}, 569-587 (2008) 

\bibitem{2012ApJ...749..108P} 
Prada Moroni, P.~G., Gennaro, M., Bono, G., Pietrzy{\'n}ski, G., Gieren, 
W., Pilecki, B., Graczyk, D., 
\& Thompson, I.~B., ApJ, \textbf{749}, 108 (2012) 

\bibitem{2009A&A...498..425K} 
Kervella, P., M{\'e}rand, A., 
\& Gallenne, A., A\&A, \textbf{498}, 425 (2009) 

\bibitem{2010ApJ...725.2392M} 
Marengo, M., Evans, N.~R., Barmby, P., et al., ApJ, \textbf{725}, 2392 
(2010) 

\bibitem{2017A&A...597A..73N} 
Nardetto, N., Poretti, E., Rainer, M., et al., 
A\&A, \textbf{597}, A73 (2017) 

\bibitem{1998ApJ...498..360S} 
Saio, H., \& Gautschy, A., ApJ, \textbf{498}, 360 (1998) 

\bibitem{2013ApJ...768L...6M} 
Marconi, M., Molinaro, R., Bono, G., et al., ApJ, \textbf{768}, L6 (2013) 

\bibitem{1969A&A.....1..142K} 
Kippenhahn, R., \& Smith, L., A\&A, \textbf{1}, 142 (1969) 

\bibitem{2011Natur.477..188M} 
Matsunaga, N., Kawadu, T., Nishiyama, S., et al., Nature, \textbf{477}, 188 
(2011) 

\bibitem{2005ApJ...621..966B} 
Bono, G., Marconi, M., Cassisi, S., Caputo, F., Gieren, W., 
\& Pietrzynski, G., ApJ, \textbf{621}, 966 (2005)

\bibitem{2013MNRAS.434.2238A} 
  Anderson, R.~I., Eyer, L., \& Mowlavi, N., MNRAS, \textbf{434}, 2238 (2013)

\bibitem{2014A&A...566A..21G} 
Georgy, C., Granada, A., Ekstr{\"o}m, S., Meynet, G., Anderson, R.~I., 
Wyttenbach, A., Eggenberger, P., 
\& Maeder, A., A\&A, \textbf{566}, A21 (2014) 

\bibitem{RIAPHD}
Anderson, R.~I., PhD thesis, Universit\'e de Gen\`eve (2013)

\bibitem{2015ApJ...807...25B} 
Brandt, T.~D., \& Huang, C.~X., ApJ, \textbf{807}, 25 (2015) 

\bibitem{2008AcA....58..163S} 
Soszynski, I., Poleski, R., Udalski, A., et al., AcA, \textbf{58}, 163 
(2008) 

%% \bibitem{2012Natur.481...55B} 
%% Beck, P.~G., Montalban, J., Kallinger, T., et al., Natur, \textbf{481}, 55 
%% (2012) 

%% \bibitem{2014ApJ...786...80G} 
%% Gieren, W., Pilecki, B., Pietrzy{\'n}ski, G., et al., ApJ, \textbf{786}, 80 
%% (2014) 

\end{thebibliography}

\end{document}